\documentclass[]{interact}

\usepackage{aas_macros}

\usepackage{xcolor}
\usepackage{epstopdf}% To incorporate .eps illustrations using PDFLaTeX, etc.
\usepackage[caption=false]{subfig}% Support for small, `sub' figures and tables

\usepackage[numbers,sort&compress]{natbib}% Citation support using natbib.sty
\bibpunct[, ]{[}{]}{,}{n}{,}{,}% Citation support using natbib.sty
\renewcommand\bibfont{\fontsize{10}{12}\selectfont}% Bibliography support using natbib.sty
\makeatletter% @ becomes a letter
\def\NAT@def@citea{\def\@citea{\NAT@separator}}% Suppress spaces between citations using natbib.sty
\makeatother% @ becomes a symbol again

\theoremstyle{plain}% Theorem-like structures provided by amsthm.sty

\theoremstyle{definition}

\theoremstyle{remark}

\newcommand{\be}{\begin{equation}}
\newcommand{\ee}{\end{equation}}

\begin{document}

%\articletype{ARTICLE TEMPLATE}% Specify the article type or omit as appropriate

\title{Axions for amateurs}

\author{
\name{David J. E. Marsh\textsuperscript{a}\thanks{CONTACT David J. E. Marsh. Email: david.j.marsh@kcl.ac.uk}}
\affil{\textsuperscript{a} Theoretical Particle Physics and Cosmology, King's College London, Strand, London, WC2R 2LS, UK}
}

\maketitle

\begin{abstract}
Axions are an increasingly popular topic in theoretical physics, and are sparking a global experimental effort. In the following I review the motivations for the existence of axions, the theories underlying them, and the methods to search for them. The target audience is an interested amateur, physics undergraduate, or scientist in another field, and so I use no complicated mathematics or advanced theoretical topics, and instead use lots of analogies.
\end{abstract}


\section{Invitation: a century of progress and problems}

We live at an extraordinary time in scientific history: never before have we known so much about the Universe, yet been so certain about our ignorance of it. %Let's begin with what we have learned in the twentieth century, starting with General Relativity (GR), and its application to cosmology.

%\subsection{Primer in cosmology}

\subsection{Newton, Maxwell, and Einstein}

In the world of Newton and Descartes three spatial coordinates, $x,y,z$, are used to describe the locations of objects, including extended objects that are imagined to be composed of infinitesimal regions of size $dx,dy,dz$. Dynamics result from differential equations describing how the coordinates of an object change with time. A useful example of an equation in classical mechanics is the Navier-Stokes equation, which describes the motion of a fluid (here assumed to be incompressible for simplicity):
\begin{equation}
\frac{\partial \mathbf{u}}{\partial t}+(\mathbf{u}\cdot \nabla)\mathbf{u}-\nu \nabla^2 \mathbf{u} = -\frac{1}{\rho}\nabla p + \mathbf{g}\, , \label{eqn:navier_normal}
\end{equation}
where $\mathbf{u}$ is the velocity of a fluid element at location $(x,y,z)$, $\nu$ is the viscocity, $\rho$ the fluid density, $p$ the pressure, and $\mathbf{g}$ an external force. The fluid velocity is a vector and its direction is specified with reference to unit vectors aligned with the coordinate axes, $\hat{e}_x,\hat{e}_y,\hat{e}_z$, with components $(u_x,u_y,u_z)$. It appears in these equations that the coordinates $x,y,z$ might be fixed, and the velocity of the fluid absolute. 

It is possible, however, to recast the Navier-Stokes equations in `tensor form' whereby we learn that there is no special role for either the coordinates $x,y,z$ themselves, or the absolute standard of rest. This form is (you can find a derivation and definitions in any book on fluid mechanics): 
\be
\frac{D \mathbf{u}}{dt} = \frac{1}{\rho}\nabla\cdot \mathbf{\sigma}+\mathbf{g}\, ,
\ee
where $D/dt$ is th convective derivative, and $\mathbf{\sigma}$ is the `stress tensor', which you can think of as a matrix $\sigma_{ij}$, which `points' in the space directions with components $\sigma_{xx},\sigma_{xy},\sigma_{xz}$, and so on. Lagrange and Hamilton also realised that we are not bound to the Cartesian $(x,y,z)$ coordinates, but that we can use `generalised coordinates' $\mathbf{q}$ (for example, we might use cylindrical polar coordinates to describe a fluid in a cylinder). The properties of the fluid are also `Galilean invariant' (fluid experiments work the same in a stationary laboratory, or in one moving at a constant speed e.g. on a train). So far, so classical.

However, despite an invariance of these equations under changes of coordinates, there still appears to be a special role for the absolute structure of the space in which events take place. For example, we find the distance, $D$, from one point in the fluid to another by squaring the vector that joins them, which in Cartesian coordinates is:
\begin{equation}
D^2 = \Delta x^2+\Delta y^2 +\Delta z^2\,.
\label{eqn:euclidean_distance}
\end{equation}
The actual distance $D$ does not change when we use different coordinates, or if we are moving relative to the fluid.

Despite being able to cast the Navier-Stokes equations in tensor form and change our coordinates, there are some things we can't do. For example, the equations are only first order in the time derivative, $\partial/\partial t$, while they are second order in spatial derivatives via the Laplacian, $\nabla^2$ (most obvious in the original form we gave in Eq.~\ref{eqn:navier_normal}). This means that we can't make changes of coordinates that mix up space and time without changing the form of the equations (a Galilean transformation with constant velocity is an exception to the rule, since only derivatives of $\mathbf{u}$ appear in Navier-Stokes in tensor form). Also, our Navier-Stokes vectors and tensors, like the fluid velocity itself $\mathbf{u}$, only `point' in the spatial directions. We don't need a `timelike' direction for $\mathbf{u}$ with a timelike unit vector $\hat{e}_t$ and component $u_t$. Similarly, the fluid stress tensor also has no component $\sigma_{tx}$ or others `pointing' in the time direction. Space and time are fundamentally different in Newtonian mechanics, with time flowing independently `from the outside', the same for all observers and all coordinate systems.

Einstein began to see the cracks in this picture thanks to the dawn of the theory of electromagnetism due to Maxwell in the late 19th century. It is said that Maxwell's theory represented both the apex, and the end, of classical physics. The familiar form for Maxwell's equations for the electric field, $\mathbf{E}$, and magnetic field $\mathbf{B}$ is:
\begin{align}
    \nabla \cdot \mathbf{E} &= \frac{\rho}{\varepsilon_0} , \\
    \nabla \cdot \mathbf{B} &= 0, \\
    \nabla \times \mathbf{E} +\frac{\partial \mathbf{B}}{\partial t}&= 0, \\
    \nabla \times \mathbf{B}- \mu_0 \varepsilon_0 \frac{\partial \mathbf{E}}{\partial t} &= \mu_0 \mathbf{J} ,
\end{align}
where $\rho$ is the free electric charge density, $\mathbf{J}$ is the electric current density, $\varepsilon_0$ is the vacuum permittivity (electric constant), and $\mu_0$ is the vacuum permeability (magnetic constant). The first thing that is different between these equations and the Navier-Stokes equations is that they are the same order in space and time derivatives. Secondly, when they are recast in tensor form the main object of concern is the Faraday tensor, $\mathbf{F}$ which we can think of as a \emph{four by four matrix} with some components \emph{in the time direction}, for example $F_{tx} = E_x$ (for the explicit form and derivation, I enjoy the presentation by Goldstein~\cite{goldstein:mechanics}). 

Maxwell's equations are equations for abstract objects known as \emph{fields} (the electric and magnetic fields) in \emph{spacetime}, in contrast to the Navier-Stokes equations that are equations for physical (tangible, touchable) fluids composed of particles, which live in space and evolve in time. When Einstein developed the concepts to describe Maxwell's equations as equations in spacetime, he was ultimately led to the conclusion that not only did he need equations to describe fields (and fluids) evolving in spacetime, but he also required equations to describe \emph{spacetime itself evolving in spacetime}: the theory of GR. The simplest example of using GR in this way is also one of the most profound: cosmology.~\footnote{For further reading on GR I recommend the introductory book by Schutz~\cite{schutz:2009} for practical purposes, while the `first track' in Misner, Thorne, and Wheeler~\cite{Misner1973} contains lots of thought experiments and intuition. For those keen to do research, I enjoy Carroll~\cite{carroll2003spacetime}.}

When we look at very distant galaxies, we see something strange. The light coming from those galaxies is very red. It isn’t caused by the light getting dimmer, or any kind of filter or gas in the way. Thanks to quantum mechanics, the light from different atoms and molecules always comes in the same colour as seen sitting next to the atom, and each colour is in a fixed ratio to the others. By lining up those colours of light seen from different galaxies, we can see that the colours are systematically shifted compared to the same atoms and molecules on Earth: all light from distant galaxies is more red than it is here on Earth. 

The only explanation for this weird fact that fits with everything else we know about the Universe, is that the light is losing energy as it travels to us. Red light has lower energy than blue light: that’s why you wear sunblock against `ultraviolet' (very blue) rays from the sun. Light is losing energy as it travels to us because the space in between those distant galaxies and us is changing: we say the Universe is expanding, but what is really happening is that distances are getting longer relative to what they were in the past. As light travels over these increasing distances, it loses energy, and so becomes redder. The fabric of spacetime itself is changing over time.

In GR the fundamental object of consideration is the \emph{spacetime metric tensor}, which tells us how to measure the distance between events in spacetime. If the Universe is expanding in a homogeneous and isotropic way (the same everywhere and in all directions), then the metric tensor gives the spacetime separation, $s^2$, in analogy to Eq.~\eqref{eqn:euclidean_distance} as:
\begin{equation}
\Delta s^2 = -c^2\Delta t^2+a(t)^2(\Delta x^2+\Delta y^2 +\Delta z^2)\,, \label{eqn:spacetime_sep_friedmann}
\end{equation}
where $a(t)$ is the `cosmic scale factor'. GR provides dynamical equations for $a(t)$ known as the Friedmann equations. The simplest Friedmann equation is:
\begin{equation}
\frac{da}{dt} = a\sqrt{\frac{8\pi G_N}{3 c^2}\rho}\, ,
\label{eqn:friedmann}
\end{equation}
where $G_N$ is Newton's gravitational constant, $c^2=1/\epsilon_0 \mu_0$ is the speed of light in vacuum (notice how this is determined by physical constants in Maxwell's equations), and $\rho$ is the energy density. This very profound equation allows us to conceive the science of the whole Universe called cosmology: the matter content of the Universe affects its evolution as given by the scale factor. Vice versa, the dynamics of $a(t)$ affects the matter density, via the continuity equation, which for a pressureless fluid reads:
\be
\frac{d\rho}{dt}+\frac{3}{a}\frac{da}{dt}\rho = 0\, .
\ee

If the energy density is dominated by a fluid with constant positive pressure, the Friedmann equation, Eq.~\eqref{eqn:friedmann}, has solutions where $a(t)\propto t^p$, i.e. the scale factor grows with time, which is consistent with the observed fact going back to Hubble that distant galaxies appear to recede from us, leading to the idea of an expanding Universe. If the Universe expands, then energy density, $\rho$, decreases with time going into the future. If we know what the Universe contains today, then we can also ask what it was doing in the past. Reversing time we would see the expanding Universe contract, and so the energy density goes up in the past. As the energy density goes up, the Universe gets hotter and hotter, eventually reaching the state of a plasma known as `the hot big bang'.

\subsection{Modern cosmology and dark matter}

GR and the theory of cosmology were developed extensively during the mid to late twentieth century. When precise cosmological measurements began to be made, and in particular of something called the `cosmic microwave background' (CMB) in the 1990's and early 2000's, cosmologists began to realise that the Universe was holding a dark secret (for a review of the development of these measurements, see Page's book~\cite{page_cosmology}). So what is the CMB? %The last piece of evidence for dark matter came at the dawn of the millennium and amounts to the most precise measurement we have of how much dark matter there is, and what some of its properties, like temperature and pressure, are. 

As we look further into space, past the galaxies that told us about the expansion of the Universe, eventually things go black - quite literally. There was a period around 13.5 billion years ago, called the `cosmic dark ages', when there were no stars in the entire visible Universe. Stars form when clouds of gas are attracted by their own gravity, and the gas gets so closely packed that a nuclear chain reaction starts. Although in the early Universe, the \emph{average} the energy density was higher (since $\rho$ goes up), it was also very smooth and flat. This is because it takes time for the inward force of gravity to overcome the expansion of the Universe and allow galaxies to form, in a process known as `hierarchical structure formation'. Before galaxies could form, during the cosmic dark ages, the gas density is too low, and those nuclear chain reactions that form stars couldn't begin.

Further back still, another strange thing happens. We stop being able to see through the Universe at all, and it becomes very bright again, but bright only in the microwave part of the electromagnetic spectrum. This is the CMB, and it is about 13.7 billion light years away from us.

Why can’t we see through the CMB? Well, why can’t we see through the Sun? The Sun is made of hydrogen and helium gas. Here on Earth, you can see through hydrogen and helium gas, but when you heat them up past about 3000 degrees, they break apart. An atom of hydrogen is one proton (positive), and one electron (negative). Put them together, and hydrogen is electrically neutral. Break them apart, and you have charged particles floating around freely. Rays of light are electromagnetic waves, and they interact with charged particles. In a soup of free protons and electrons (a plasma, like the Sun) light gets bounced around so much it can’t get through. If you heat hydrogen up to a 3000 degree plasma, it becomes opaque: light can’t get through.

Scientists in the mid twentieth century knew that the Universe was expanding, and they also knew it contained a lot of hydrogen gas. They ran their equations backwards and concluded that at some point in the past the Universe would have been so dense that it would turn into one gigantic star: a plasma of hot gas and electromagnetic waves, that we wouldn’t be able to see through. These scientists also predicted, by taking the 3000 degrees it takes to break apart hydrogen, and working out how far back in time you have to go until the Universe is that dense and hot, and then working out how much energy the light would lose travelling to us from that time, that this gigantic star should be visible to us now, as microwave radiation at about -270 degrees,  covering the whole sky. And that is exactly what the CMB is. We are viewing the Universe as the inside of a gigantic star.

This is where things get really interesting.

Because we can’t see beyond the CMB, we see it as a single surface of microwave radiation, coming at us from all directions, all the time. This single surface is like a photograph of the state of the Universe 13.7 billion years ago. By looking closely at it, and trying to understand it and how it connects to the Universe we see today, cosmologists can learn about the past, present, and the future, of the Universe.%, using the predictive power of physics and mathematics.

The cosmic microwave background is radiation left over from the big bang that we observe filling the night sky at a cool -270 degrees Celsius: a picture of the Universe as it was 13.7 billion years ago. Using satellites and ground-based radio telescopes, astronomers can observe small differences in this temperature across the sky, differences at one part in 100 thousand, and use this to map out the distribution of matter in the very early Universe. Using GR, specifically perturbation theory based on the Friedmann equations, cosmologists can work out how the Universe evolved from 13.7 billion years ago until today. This allows us do two things: 
\begin{enumerate}
\item Find out the matter content and initial conditions required to correctly describe observations of the CMB.
\item Evolve that Universe further forward and compare it to the actual Universe observed today.
\end{enumerate}

`The actual Universe observed today' is shorthand for observations made by telescopes that map the locations of 100s of millions of different galaxies, which are seen to cluster together in a grand structure called the cosmic web (two of the many instruments to make these measurements are the Sloan Digital Sky Survey~\cite{2013AJ....145...10D} and the Dark Energy Survey~\cite{DES:2020aks}).  %The next leap forward in these measurements is about to begin with the launch of the European space agency’s Euclid satellite earlier this summer, something I had a small part in during my PhD back in 2011.

Now the critical part. The Universe we observe in the CMB and the cosmic web of galaxies only evolve consistently one into the other if we add an extra ingredient into the equations of GR. We must add a new form of matter in the form of a (very) cold, (almost) pressureless fluid. This fluid is called dark matter (DM), and its total mass is measured to a precision of 1\% using the CMB. The idea of this interplay between measurements and theory that is used to determine the DM density consistently between the early Universe CMB, and the late Universe cosmic web is known as hierarchical structure formation and underpins modern cosmology~\cite{dodelson:2003}. 

The inferred average density of DM from measurements of the CMB by the \emph{Planck} satellite is~\cite{Planck:2018vyg}:
\begin{equation}
\rho_{\rm DM} = 2.3\times 10^{-27}\text{ kg m}^{-3}\, , \label{eqn:cmb_relic_density}
\end{equation}
roughly equivalent to the mass of a single hydrogen atom in a cubic metre. This number might seem incredibly tiny, and it is: space is very empty. This should be compared to the average density of ordinary matter (mostly in the form of hydrogen gas), which can be inferred in a similar way:
\begin{equation}
\rho_{\rm matter}=4.2\times 10^{-28}\text{ kg m}^{-3}\, . \label{eqn:baryon_relic_density}
\end{equation}
There is about five times more DM than ordinary matter in the Universe. To give a sense for how such a tiny number can be measured, we know that the average density of DM in the Milky Way is about 10000 times larger than Eq.~\eqref{eqn:cmb_relic_density}. If you multiply this by the volume of the Milky Way, very roughly $2\times 10^{17}$ cubic light years, then we get a mass more than a million million times the mass of the Sun, which is more than the total mass of visible stars in the Milky Way.

It is often said that we don’t know what DM is, but that is not really true. The macroscopic theory of a cold, pressureless fluid works perfectly well to describe cosmological measurements. Furthermore, the possibilities for microscopic theories are severely restricted: they have to behave like the cold fluid necessary to explain the structure of the CMB, and whatever makes up that cold fluid had to be present in the very early Universe in the primordial plasma.~\footnote{We focused on evidence for DM from the CMB because it is \emph{impossible} to explain the CMB any other way. Modifying gravity doesn't work without also introducing new dark degrees of freedom, i.e. without introducing DM.} The mystery is: what is the microscopic theory of dark matter? Axions provide one possible answer, and one that is gaining popularity among theorists and experimentalists.

\section{What is an axion?}

\subsection{One particle, two mysteries}

\subsubsection{Strong-$CP$ problem}
%%%%%%%%%%%%%%%%%%%%%%
\begin{figure}
    \centering
    \includegraphics[width=0.6\textwidth]{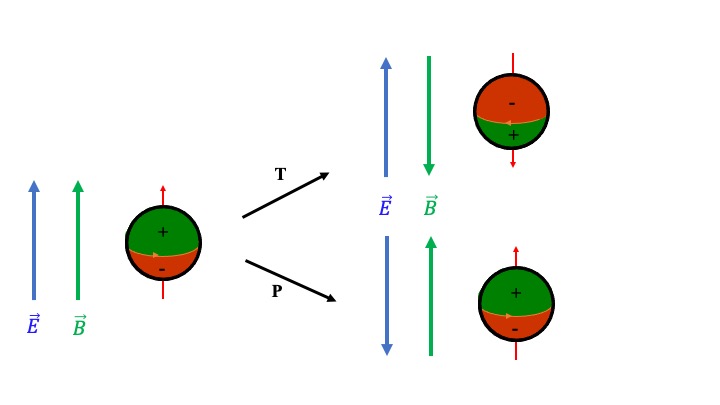}
    \caption{The effect of time, $T$, and parity, $P$, transformations on an electric or magnetic dipole moments (which must \emph{both} point in the same direction as the spin given by the red arrow). We can work out the effect by considering that $CPT$ must be conserved, and by the action of $T$ and $P$ and on the applied electric field and on a magnetic field (recall that $\mathbf{B}$ can be thought of as sourced by a current, so $T$ changes its direction). (Reproduced from Ref.~\cite{Chadha-Day:2021szb})}
    \label{fig:EDM}
\end{figure}
%%%%%%%%%%%%%%%%%%%%%%

The story of axions begins not with DM, but with another mystery concerning particle physics.The Standard Model of particle physics contains electrons, and their heavier siblings, mu and tau, and the lighter cousins called neutrinos. Then we have quarks: up, down strange, charm, top and bottom, in three colours each, red, blue, and green. For every one of these particles, there is an anti-particle as well. Next there are force carriers, called bosons: the photon of electromagnetism, the gluon that holds protons and other things together with the so-called strong nuclear force, and the $W$ and $Z$ bosons responsible for radioactivity (the weak nuclear force). Lastly there is the Higgs boson, responsible for giving mass to the other particles. 

There are three `discrete' symmetries commonly considered in reference to the Standard Model: charge conjugation (denoted by $C$), `parity reversal' or `inversion symmetry'  (denoted by $P$) and time reversal, $T$. Each symmetry, denote it $S$, acts on a particle state, $\psi$ as:
\be
S\psi = (\pm 1) \psi\, ,
\ee
if the particle state is even ($+1$) or odd ($-1$) under the symmetry. It is believed that the combination $CPT$ should always leave any state even, i.e. doing all of them in turn returns $(+1)$. Radioactive decays mediated by the $W$, $Z$ and photon (unified as the `electroweak force') treat particles and anti-particles differently. Specifically, the weak interactions violate the combination of symmetries $CP$: if states before an interaction are $CP$ even, then afterwards the states that come out are $CP$ odd. This symmetry violation was first observed in the 1960's in decays of $K$-mesons~\cite{Christenson:1964fg}, and is now known to occur in many other situations involving the electroweak force. Interactions mediated by the gluons of the strong force, however, are more democratic, and do not distinguish between particle and anti-particle. The mystery of why \emph{for the same particles} the electroweak force violates $CP$, yet the strong force conserves it is known as the `strong-$CP$ problem'.

Before we get into some physics of the  strong-$CP$ problem, it is useful to begin with an analogy, which I have taken from Sikivie~\cite{Sikivie:1995pz}. Let's imagine we want to play pool or snooker, or do anything that requires a very flat surface. The world around us, however, is very much not flat. How can you make a very flat surface in a room with wonky floors? The answer is to attach a pendulum or other large weight to the surface along with a freely rotating cantilever arm. The weight of the pendulum will pull the arm perfectly vertical, and the attached table will be perfectly flat, like a spirit level. To make this all stable, you will probably want some vibration isolating springs. When you first install this pendulum and spring-loaded table, there will be a little bit of wobble caused by the installation, but thanks to our vibration isolating springs this wobble will go away with time, depositing energy into the springs. If you had a very keen eye, you might be able to work out the pendulum and springs mechanism was there by the response of the balls moving on the table and bouncing off the cushions.

Now let's get back to physics. If the strong force violated $CP$ like the weak force then strongly interacting particles, like the neutron, $n$, should have $CP n = -n$. Because the neutron should be even under $CPT$, then if $CPn=-n$ then $T n=-n$ too, i.e. $CP$ violation implies $T$ violation. This should help you understand how we can measure whether the strong force violates $CP$ or not, as we now explain. 

Consider an `electric dipole moment' (EDM), $\mathbf{d}$, which means that there is an energy associated to application of an electric field, $\mathbf{E}$ to a particle. The Hamiltonian is:
\be
H = -\mathbf{d}\cdot\mathbf{E}\, . \label{eqn:nEDM_H}
\ee
In quantum mechanics the Hamiltonian measures energy, so this equation tells you the energy of the system i.e. there is a lower energy if the EDM, $\mathbf{d}$, aligns with the applied $\mathbf{E}$ field. The neutron is known to have a magnetic dipole moment (MDM), $\mu$, for which the Hamiltonian is similarly, $H=-\mu\cdot\mathbf{B}$. Now consider what $P$ and $T$ symmetries do to an EDM and an MDM, by their action on $\mathbf{E}$ and $\mathbf{B}$, illustrated in Fig.~\ref{fig:EDM}. With some thought, this diagram should convince you that the existence of a neutron EDM (or an EDM for any strongly interacting particle) would violate $CP$, because it separately behaves differently from the MDM under $P$ and $T$. 

It is, however, an experimental fact that the neutron EDM is undetectably small (i.e. zero within very small experimental errors). By exposing a large number of neutrons to electric and magnetic fields, experimentalists can try and measure whether there is any relative energy splitting caused by the $\mathbf{E}$ field and measure the Hamiltonian co-efficient $\mathbf{d}$ in Eq.~\eqref{eqn:nEDM_H}. The value of $\mathbf{d}$ is found to be consistent with zero to very high precision~\cite{Pendlebury:2015lrz}. Thus, the strong force must in fact conserve $CP$, as we stated above.

In the theory of the strong interactions, known as \emph{quantum chromodynamics} (QCD), the amount of $CP$ violation is expressed by a \emph{dimensionless} number, $\theta$, which can be visualised classically as the base angle between the three up and down quarks in the neutron arranged in an isosceles triangle. The size of this angle is a \emph{constant} in the theory, and determines the neutron EDM as $\mathbf{d}\propto \theta$.~\footnote{The constant of proportionality can be estimated by dimensional analysis. An EDM has units charge times distance. The charge we have to play with is the quark charge, $e/3$, and the distance is the size of the neutron, $10^{-15}\text{ m}$. So we estimate the constant as the product of these numbers, about $3\times 10^{-14}\,e\text{m}$. The value of the neutron EDM computed using quantum field theory~\cite{Crewther:1979pi} is $d=5\times 10^{-14}\theta\, e\text{m}$: very close to our naive estimate.} The observed smallness of $\mathbf{d}$ implies $\theta\approx 10^{-10}$, i.e. in our classical triangle picture the quarks all lie on an almost perfect straight line. The strong $CP$ problem is that $\theta$ has two unrelated contributions: from the electroweak force (that violates $CP$) and from the strong force. A small value of $\theta$ implies a delicate cancellation between two numbers. 

\subsubsection{Axions to the rescue}

%%%%%%%%%%%%%%%%%%%%%%
\begin{figure}
    \centering
    \includegraphics[width=0.6\textwidth]{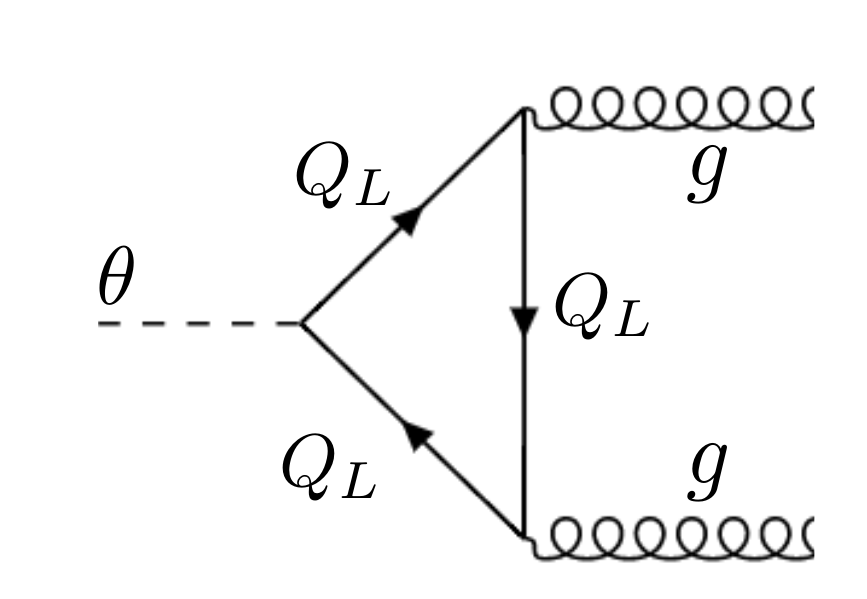}
    \caption{Feynman diagram that couples the axion field, $\theta$, to the gluons, $g$ via quark, $Q$. This diagram, if it has opposite sign for left and right handed quarks, will ensure that $\theta$ is the same $\theta$ responsible for the neutron EDM.}
    \label{fig:colour_anomaly}
\end{figure}
%%%%%%%%%%%%%%%%%%%%%%

In the 1970's, theoretical physicists realised this situation of having and yet not having $CP$ symmetry in the Standard Model was a problem, and set about trying to explain why it might be. The solution came in 1977 thanks to Peccei and Quinn~\cite{Peccei:1977hh}. Peccei and Quinn realised that if they could `promote' $\theta$ from a constant to field, which is not a constant but varies throughout spacetime (like the fields of electromagentism that we met above), and if they could also make $\theta=0$ somehow energetically favourable, then the strong-$CP$ problem would solve itself dynamically. We will only have space here to outline Peccei and Quinn's method for achieving this trickery.

The beginning of the solution is to answer the question: `when is a field an angle?', and the answer is `when it is a complex number'. A complex number $z$ is first introduced to us as $z=x+iy$ with $x$ the real axis and $y$ the imaginary axis. Since the complex numbers can thus be thought of as coordinates on the `Argand plane', with $x$ and $y$ the Cartesian coordinates, we can also use other coordinates, in particular plane polar coordinates. In these coordinates, $z=Re^{i\theta}$ and $x=R\cos \theta$, $y=R\sin \theta$ (do not confuse these $x$ and $y$ with $x$ and $y$ coordinates in spacetime!). Peccei and Quinn introduced a new field $\Phi$ to the Standard Model, which takes on a complex value everywhere in spacetime, i.e. in polar coordinates in the complex plane, and Cartesian coordinates in spacetime:
\be
\Phi = R(t,\mathbf{x})\exp [ i \theta(t,\mathbf{x})]\, . \label{eqn:PQ_polar}
\ee

Now we have identified the right field content, we need to try and make the $\theta$ in Eq.~\eqref{eqn:PQ_polar} the same $\theta$ responsible for the neutron EDM. As Feynman explains it~\cite{feynman_quantum_1998}, to every field there is a particle, and particle physicists compute the effects of interactions between particles by drawing squiggly little diagrams and adding up the values of stopwatches carried round by these particles as they follow the lines on the diagrams. 

The particular squiggly diagram that allowed Peccei and Quinn to make the $\theta$ in Eq.~\eqref{eqn:PQ_polar} the same $\theta$ responsible for the neutron EDM is shown in Fig.~\ref{fig:colour_anomaly}. The interaction we need is one between $\theta$ (which we will now start calling the `axion particle'~\footnote{The name `axion' is due to Frank Wilczek. It was Weinberg and Wilczek who, independently later in 1977 (published in 1978)~\cite{Weinberg:1977ma,Wilczek:1977pj} first realised that Peccei and Quinn's theory predicted the existence of a particle, and computed its mass. Wilczek coined the phrase `axion' after the American detergent. The `axi' comes from the left/right handed necessity of the interaction between axions and quarks, which physicists call `axial', while the `on' just sounds like a particle name (think `boson', `neutron' etc.). The axion `cleans up the mess' of the strong-$CP$ problem. Weinberg's name for the particle was the `Higglet', since it is a bit like a Higgs boson, only lighter.}) and a quark, $Q$. The interaction shown is between $\theta$ and `left handed' quarks only, i.e. those where the spin angular momentum points opposite to the linear momentum. For Peccei and Quinn there must also be an equivalent diagram for right handed quarks, but with an opposite sign. If we are allowed to draw these particular diagrams then when we add up all the little stopwatches we find that the axion field, $\theta$, couples to gluons, $g$, in just the right way that the value of the neutron EDM is determined, at every point in spacetime, by the value of the axion field, $\theta$.~\footnote{The actual computation requires a graduate course in quantum field theory. You can find it in these references~\cite{Srednicki:1019751,Zee:706825}.}

\begin{figure}
    \centering
    \includegraphics[width=0.6\textwidth]{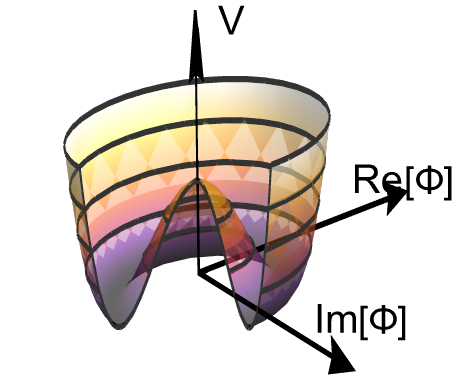}
    \caption{`Wine bottle' potential for the complex field $\Phi$ (the angular direction is opened up for visualisation). The potential is minimised at a fixed value of $R=|\Phi|=v$. We do not show the small perturbation $\Delta V\sim \theta^2$ that `tilts the wine bottle' and minimises it at a fixed angle $\theta=0$. Think about how the dreggs in a real wine bottle accumulate in the bottom of the punt when the wine is poured, and you are visualising the axion field reaching the minimum of its potential.}
    \label{fig:wine_bottle}
\end{figure}

The final ingredient in the Peccei-Quinn recipe is to make $\theta=0$ energetically favourable. Making a particular field value energetically favourable means writing down a potential energy that depends on the value of the field, $V = V(\Phi)$, which has a minimum at the desired value. Firstly, the potential $V$ should very strongly favour $R = v\gg 0$, because the polar coordinate representation Eq.~\eqref{eqn:PQ_polar} is only valid away from the origin of the complex plane: at the origin $\theta$ is not defined. This part of the problem is solved with a `wine bottle' type potential, like the Higgs has in the Standard Model, as pictured in Fig.~\ref{fig:wine_bottle}. The potential should also favour $\theta=0$, to solve the strong $CP$ problem and set the neutron EDM to zero, the simplest option being $V(\theta)\sim \theta^2$, i.e. a harmonic potential, which in quantum field theory corresponds to a mass for the axion particle. The bit that favours $\theta=0$ should not be too big to mess with the definition of polar coordinates: it should be a perturbation to the leading wine bottle piece. This part of the potential comes for free due to the strong force itself and the way that the axion interacts with mesons and the `topology of the QFT vacuum' (related to large amplitude but short-lived fluctuations of the gluon fields called instantons).

Because the axion mass comes for free from the strong force itself, it turns out that the only free parameter in this model is the location in field space where $V$ is minimised in the $R$ direction, i.e. the value of what we called $v$, which determines something called the `axion decay constant', denoted $f_a$. The axion mass, and the strengths of its interactions with all the particles of the Standard Model are determined by the value of $f_a$. The best current computation of the axion mass (at low temperatures) is~\cite{GrillidiCortona:2015jxo}:
\be
m_a = 5.7(1) \, \, \mu\text{eV}\left( \frac{10^{12}\text{ GeV}}{f_a}\right)\, .
\ee
Notice that if $f_a$ takes the very large reference value given here, the axion is extremely light (c.f. the electron, which has a mass $m_e=511\text{ keV}$). Furthermore, if $f_a$ increases, the axion mass goes down, and so does the strength of all its interactions: everything goes \emph{inversely} with $f_a$. We do not know the axion mass, and the task of discovering the axion is to measure its mass experimentally. 

%%%%%%%%%%%%%%%%%%%%%%
%%%%%%%%%%%%%%%%%%%%%%

\subsubsection{Axion dark matter}

Still in the 1970's, particle physicists realised that if the unknown constant in the axion model, $f_a$, was set to be close in value to a similar parameter in the Standard Model, the Higgs vacuum expectation value $v_H= 246\text{ GeV}$ (the value it had to take in the original models), then the axion would have already shown up in their experiments. Was the model of Peccei, Quinn, Weinberg, and Wilczek dead? Shortly afterwards, between 1979 and 1981, eight different physicists working in four different groups on either side of the iron curtain cooked up two different variations on the model~\cite{Kim:1979if,SHIFMAN1980493,Dine:1981rt,Zhitnitsky:1980tq}. In these variations, $f_a$ was allowed to be very large, making the axion `invisible' to the particle physicists' experiments (recall: we said above that interaction strengths are inverse with $f_a$). 

The next step in the story came in 1983 when three different groups started to think about what happened to this invisible axion in the early Universe~\cite{1983PhLB..120..137D,1983PhLB..120..127P,1983PhLB..120..133A}. They did this by considering the wave equation that describes the dynamics of the axion field in an expanding Universe with scale factor $a$ that we introduced above. We will study this equation, and how it produces axion DM in Section~\ref{sec:waves_cosmo}. In short, the early Universe naturally provides conditions that leave the axion displaced from the minimum energy value at $\theta=0$ (the Universe imparts some energy to the axion field). Relaxation from the displaced value to zero leads to damped oscillations in the axion field, and it is the energy density carried by these oscillations that can explain DM.

The model for axion DM introduces a new parameter, the initial value of $\theta$, and axions are capable of explaining all, or none, of the cosmic DM for a wide range of values for the particle mass, making the experimental task of finding axions very difficult. Nonetheless, the emergence of axions as a DM candidate is one of the main factors that drives interest in them today.

\subsection{Many dimensions, many axions}

The origin of axions as a solution to the strong-$CP$ problem turns out to be just one place in which they crop up in theoretical physics; another one is \emph{string theory}. 

Trying to marry the two pioneering successes of twentieth century physics, quantum mechanics and general relativity, turns out to be immensely difficult. Quantum mechanics wants everything to jump around and undergo zero-point fluctuations and for all fields to be quantised as point-like particles. General relativity wants the fabric of spacetime to be a dynamical field called the metric. If we quantise the metric as a particle called the graviton, and let it fluctuate, then all hell breaks loose on small scales and at high energies. Spacetime fluctuates and produces singularities and black holes with ever growing number, and eventually the equations just give infinity in answer to any question. String theory is a theory of \emph{quantum gravity} that replaces particles with extended objects known as strings. Because the strings have a finite length, then the short distance problems of particle quantum gravity are avoided.

General relativity predicts lots of things, but there are two things it is silent on. The first is the number of dimensions of spacetime. GR works just fine in two, three, four, or 104 dimensions. Any more than four would seem silly, but as Kaluza and Klein (and also Einstein) famously showed, you can easily hide extra dimensions, and you may even get something desirable from doing so, in terms of an ability to describe GR and electrodynamics with the same theory.~\footnote{You can read about Kaluza-Klein theory on Wikipedia.} There could be extra dimensions of spacetime and GR gives us no guidance: experimentally all we know is that if extra dimensions exist, then they must be smaller than around a micron~\cite{pdg}. Secondly, GR does not specify the \emph{topology} of spacetime. That is, if spacetime on the largest scales is closed on itself, like a ball, or a more complex shape like a torus (doughnut) or Riemann surface (multi-holed doughnut), GR does not give an answer for a dynamics of why, because changes of topology are impossible in Einstein's theory. It is an experimental question also what the topology of the four large dimensions we observe is. Cosmologists look for evidence of cosmic topology, and currently we know that if the Universe is a doughnut, or a related topology, then the size of the `fundamental domain' is larger than 98\% times the size of the visible Universe~\cite{COMPACT:2022nsu}. If there are tiny extra dimensions, their topology is virtually unconstrained.

Quantum field theory, the most fundamental theory of quantum mechanics we have on which the Standard Model is based, also fails to make predictions about certain things. QFT tells us very little about what the particle content of the Standard Model should be. We must find out experimentally which particles exist. The famous saying of Rabi regarding the muon `who ordered that?' is now more rightly `who ordered all of this?'. In particular, it is a mystery why there should be three generations in the Standard Model corresponding to the heavier siblings of the electron and quarks. Even invoking the anthropic principle, that the laws of physics should be complex enough to allow life, to explain some complexity, then the existence of the weak nuclear force seems mysterious.

String theory provides answers on all of the above questions. String theory is such a rigorous mathematical structure that it breaks down if its ingredients are changed. String theory predicts that there are ten dimensions of spacetime.~\footnote{In the weakly coupled limit. In the strong coupling limit, degrees of freedom on the strings reorganise themselves into an `emergent' 11th dimension or even a 12th `half' dimension in so-called M-theory and F-theory.} String theory also predicts that there are only strings of one type (although there are different `dual' descriptions), and all the particles we observe are related to their different modes of vibration. String theory does not predict the topology of the extra dimensions (although unlike GR it does allow for some changes of topology), however the particle content we observe in the four large dimensions is intimately related to topology. It is here that axions appear in the story.

At low energies, and in ten dimensions, the dynamics of strings in string theory reduces to what is called \emph{supergravity}. Supergravity is a theory of the metric (like GR), fermions (like the known particles including electrons), and a host of other fields known as $p$-form fields. A $p$-form field is like the four-dimensional electromagnetic vector potential, $A=(\varphi,\mathbf{A})$ (where $\varphi$ is the electrostatic potential, so that $\mathbf{E}=\nabla\varphi$, and $\mathbf{B}=\nabla\times\mathbf{A}$), but a $p$-form is a rank-$p$ anti-symmetric tensor with $10^p$ components (not all of them independent). So if $p=1$, think of a vector with 10 entries, if $p=2$ think of a matrix with 100 entries, if $p=3$ think of a cube of fields with 1000 entries, and so on.

For the metric, let's first imagine we simply have five dimensions: our usual four plus a circle. The equivalent to Eq.~\eqref{eqn:spacetime_sep_friedmann} is
\begin{equation}
\Delta s^2 = -c^2\Delta t^2+a(t)^2(\Delta x^2+\Delta y^2 +\Delta z^2)+\mathcal{R}(t,\mathbf{x},\theta)^2\Delta\theta^2\,, \label{eqn:spacetime_sep_friedmann_s1}
\end{equation}
where $\theta$ is the coordinate around the circle and $\mathcal{R}$ is the radius of the circle (which can change from place to place as we move around, if the circle is bent). Einstein's equations give rise to equations of motion for the circle radius, just like the Friedmann equation gave us an equation for $a$ in Eq.~\eqref{eqn:friedmann}. How do we interpret this? An extra dimension, if small enough, would be invisible to us, just like a tightrope walker cannot move around the thickness of the tightrope, but an ant can. So if $\mathcal{R}$ is small, we won't notice immediately by walking in the fifth dimension. Thanks to GR, however, $\mathcal{R}$ is also dynamical and interacts with, for example, the scale factor $a$ and all other fields via gravity. To make things realistic there needs to be an extra energy penalty to fix $\mathcal{R}$ to a small but non-zero value and hide its dynamics from us so we don't notice. In string theory this problem is called `moduli stabilisation' and it can be done using physics similar to what we discussed for the axion potential above. Fluctuations of $\mathcal{R}$ with this potential energy behave like an extra heavy particle called a modulus from the four dimensional perspective.

Now for the $p$-form fields. Well, being like the electromagnetic vector potential, the $p$-form field also defines $p$-form electric and magnetic fields, and these obey equations just like Maxwell's equations, which also have propagating wavelike solutions. In the example of the extra dimensional circle and a 1-form just like the Maxwell field, we would need just a single field $\phi$ to describe the dynamics of the energy stored in magnetic flux wrapped around the circle as viewed from a four dimensional perspective (this example is treated in some detail in the lecture notes Ref.~\cite{Reece:2023czb}). The field $\phi$ is known as an axion, since it has the same $CP$ properties as the axion from the strong-$CP$ problem (inherited from the parity and time reversal properties of the $p$-form electric and magnetic fields). This axion does not have to couple to gluons, and might not solve the strong-$CP$ problem, although similar axions can.

String theory has six extra dimensions, not just one as in our simple circle example, and a whole host of different $p$-form fields. The possibilities for the topology are (possibly) endless, but for semi-realistic cases giving rise to the Standard Model in four dimensions are thought to be given by a topological object called a Calabi-Yau manifold, and all such topologies will have axions~\cite{Witten:1984dg,Candelas:1985en}. It is known how to construct huge numbers of different Calabi-Yaus. The number of axions (giving the dynamics of $p$-form energy flux in the extra dimensions) and moduli (giving the dynamics of the extra dimensions themselves) arising in the four dimensional theory from such a complex extra dimensional space can be large, in many cases into the hundreds, opening up a massive playground for phenomenology known as the `string axiverse'~\cite{Conlon:2006tq,Svrcek:2006yi,Arvanitaki:2009fg}. Very recently, it has been possible to also compute the masses and decay constants of the axions across large parts of the string theory landscape, allowing even to falsify some fraction of string theory models~\cite{Mehta:2021pwf,Demirtas:2021gsq}, a task that has been declared in the past by detractors of string theory to be impossible, yet axions hold the key.

\subsection{What's in a name?}

Before we carry on, a point on nomenclature. We met the axion first as a solution of the strong-$CP$ problem in quantum chromodynamics (QCD). We have also met axions in string theory. There are yet more realisations of axions in other areas of particle physics. It is common nomenclature to call all of these particles `axions', and call the axion that solves the strong-$CP$ problem (which, incidentally, is also one of the string theory axions) `the QCD axion'. We adopt this nomenclature from hereon. Elsewhere you will see authors use the convention that only the axion that solves the strong-$CP$ problem is called `axion', and all the others called `axion-like particles' (ALPs). 

\section{The physics of axions}

The way we have so far described them, axions might seem very complicated. They are fields in the complex plane coupled to gluons by quantum loops that control the neutron electric dipole moment. They are energy stored in $p$-form electric and magnetic fields threading extra dimensions of spacetime. At the level of effective field theory, however, many aspects of axion physics are remarkably simple. We only have space here to discuss two aspects of axion physics. The first is the details of how axions behave in cosmology and solve the problem of DM. The second is one way in which axions might interact with ordinary matter and allow us to search for them.

\subsection{Axion waves in the cosmos} \label{sec:waves_cosmo}

%%%%%%%%%%%%%%%%%%%%%%%
\begin{figure}
    \centering
    \includegraphics[width=0.6\textwidth]{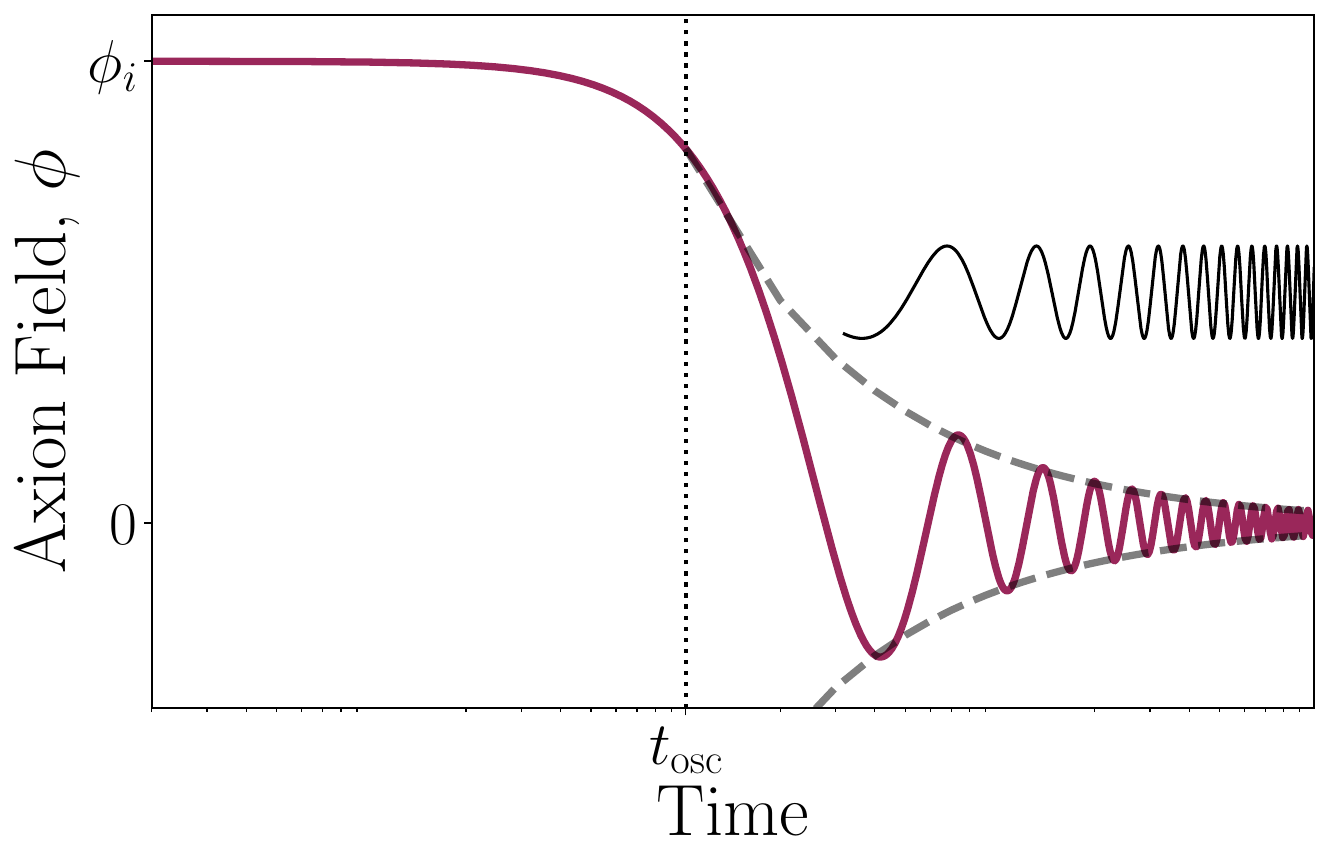}
    \caption{Numerical solution of Eq.~\eqref{eqn:KG_homo}, on a logarithmic time axis. Illustrated are the analytically estimated value of $t_{\rm osc}$ (vertical dotted line), the damping envelope (dashed line), and the undamped cosine solution (thin black line, offset for visibility). (Reproduced from Ref.~\cite{DM_book}, published by Princeton University Press, forthcoming)}
    \label{fig:axion_field}
\end{figure}
%%%%%%%%%%%%%%%%%%%%%%

We will now work with $\phi=f_a\theta$, which is the `canonically normalised axion field'. The wave equation for $\phi$ is:
\be
\frac{1}{c^2}\frac{\partial^2\phi}{\partial t^2}+\frac{3}{c^2 a}\frac{da}{dt}\frac{\partial\phi}{\partial t}-\frac{1}{a^2}\nabla^2\phi +\frac{m_a^2 c^2}{\hbar^2} \phi = 0\, .
\ee

I want you to notice a number of things about this equation. Firstly, it is a wave equation for a \emph{classical} field $\phi$, just like the wave equations we can derive for $\mathbf{E}$ or $\mathbf{B}$ from Maxwell's equations. It therefore has solutions that are propagating waves, if there is a source, just like Hertz's radio waves. The quantisation of these classical waves gives us axion particles. Secondly, this equation is equal in derivative order in space and in time, that means it is fully consistent with relativity and a field in spacetime (unlike the Navier-Stokes equations which are only an effective description of a fluid, valid above the scale of the atoms making it up). Thirdly, like Maxwell's equations, this equation is linear, but that is only because I have made the simplifying assumption to drop interaction terms between the axion and other things. It turns out that these interaction terms are very small if $f_a$ is very large, and we can neglect them at a first pass. We will return to axion interactions later. Lastly, we are still only treating the Universe itself as homogeneous, using only the cosmic scale factor $a$. Rest assured we can also include gravitational interactions of the axion with ordinary matter in perturbation theory.

Now we will make another simplifying assumption. We will assume that the axion field itself is uniform/homogeneous, i.e. it has no spatial variations, so that $\nabla\phi = 0$ and $\nabla^2\phi=0$. This must be true on average, because the Universe itself is homogeneous on average, and again corrections to this assumption can be included at a cost in mathematical complexity but without change to the basic physical picture. The equation for the axion field is now: 
\be
\frac{d^2\phi}{d t^2}+3\frac{da}{dt}\frac{d\phi}{d t} +\frac{m_a^2 c^4}{\hbar^2} \phi = 0\, . \label{eqn:KG_homo}
\ee

This wave equation should be very familiar. It is the equation of motion for a damped harmonic oscillator. The frequency is given by $\omega_a = E_a/\hbar = m_a c^2/\hbar$. The damping co-efficient is given by $\gamma = 3(da/dt)/a$: the time varying expansion of the Universe acts as a friction term on axion waves. At early times, the expansion of the Universe is fast, and if $(da/dt)/a\gg (m_a c^2)/\hbar$ the axion field is overdamped and keeps an almost constant value over time. At later times, when the expansion slows, and $(da/dt)/a\ll (m_a c^2)/\hbar$ the axion field will undergo damped oscillations. The numerical solution of this equation using a realistic model for the expansion of the Universe is shown in Fig.~\ref{fig:axion_field}, along with parts of the solution that can be estimated analytically (for a more detailed treatment, see Ref.~\cite{Marsh:2015xka}).

If the axion field starts off with some initial value $\phi_i>0$, then it will begin to oscillate at some time $t_{\rm osc}$ when the expansion of the Universe is slow enough that the damping becomes small. The oscillations of the axion field carries energy density given by:
\be
\rho = \frac{1}{2}\left(\frac{d\phi}{dt}\right)^2+\frac{1}{2}m_a^2 c^2 \phi^2\, .
\ee
\emph{The fact that the oscillations of the axion field carry energy density, yet interact very weakly with ordinary matter, makes these axion waves a candidate to explain DM.} The axion DM is effectively created at the time when the field begins to oscillate, which for large enough axion mass, $m>10^{-24}\text{ eV}$, happens sufficiently early to match the CMB and the cosmic web, and the effective description is furthermore of a cold and pressureless gravitating fluid~\cite{Hlozek:2014lca}.

There is one additional free parameter in the model now, besides $f_a$, and that is the initial value of the axion field, $\phi_i$, which we can alternatively parameterise as $\theta_i$, which measures how far the strong force vacuum was from conserving $CP$ in the very early Universe (or in the string case, the initial amount of $p$-form flux). A non-zero, albeit random, value for $\theta_i$ is naturally provided by a process called spontaneous symmetry breaking. Taking into account that, due to the dynamics of the strong nuclear fore and the formation of protons, the `real' QCD axion (as opposed to our toy model in Eq.~\ref{eqn:KG_homo}) has a time dependent mass, the relic density parameter for QCD axion DM can be expressed in terms of these two free parameters as:
\be
\rho_a = 2.3\times 10^{-27}\text{ kg m}^{-3} \left(\frac{f_a}{10^{12}\text{ GeV}}\right)^{7/6}\left(\frac{\theta_i}{0.53}\right)^{2}\, ,
\ee
where I have chosen reference values of $f_a$ and $\theta_i$ (in radians) that reproduce the observed value given in Eq.~\eqref{eqn:cmb_relic_density}.

\subsection{Axion electrodynamics}

Axions are odd under the $CP$ symmetry: $CP\phi = -\phi$. Hamiltonians are an energy, and so overall are even: $CP H = H$. If we want to couple the axion to the Maxwell fields in the Hamiltonian, the axion can only couple to a $CP$ odd combination of the fields, and the simplest such configuration is given by $\mathbf{E}\cdot\mathbf{B}$ (which you can remind yourself is $CP$ odd with reference to Fig.~\ref{fig:EDM}). Thus we can guess that the Hamiltonian for axions coupled the Maxwell fields is:
\be
H_{\rm int} = \sqrt{\frac{\epsilon_0}{\mu_0}}g \int \phi \mathbf{E}\cdot\mathbf{B}\, dV\, , \label{eqn:em_interaction}
\ee
where the factors of $\epsilon_0$ and $\mu_0$ are chosen by convention so that $g$ is an unknown coupling constant with units inverse energy (so that particle physicists measure $g$ in GeV$^{-1}$), and the integral is taken over all space in order to get the correct units of energy for the Hamiltonian itself. This Hamiltonian tells us that there is an energy cost associated to regions of space with non-zero axion field and $\mathbf{E}\cdot\mathbf{B}$. In any region of space with non-zero $\mathbf{E}\cdot\mathbf{B}$  external source (e.g. provided by an experimentalist), the energy can be lowered by the axion field taking a negative value in response. Thus a static $\mathbf{B}$ field parallel to an oscillating $\mathbf{E}$ acts as a source of axion waves as the axion field changes sign in response to the changing $\mathbf{E}$ in order to keep the energy at its minimum. Alternatively, in any region with non-zero axion field and non-zero either $\mathbf{E}$ or $\mathbf{B}$, then the energy can be lowered by creating an antiparallel $\mathbf{B}$ or $\mathbf{E}$ field. In a region of space in which the axion oscillates, a static $\mathbf{B}$ or $\mathbf{E}$ field will source an oscillating version of the other field: as if from a Hertz-like oscillating charge density the axion would source radio waves.

Now let's see this in equations. Given a Hamiltonian, we can construct an action (in this case simply by integrating over time), and from an action, via a variational principle, one can derive equations of motion (for this derivation of the ordinary Maxwell's equations in classical mechanics, see Ref.~\cite{goldstein:mechanics}). In this way, we find how this interaction Hamiltonian modifies Maxwell's equations, and the equation of motion of the axion itself, Eq.~\eqref{eqn:KG_homo}. We will just look at the modified Maxwell's equations, which read:
\begin{align}
    \nabla \cdot \mathbf{E} &= \frac{\rho}{\varepsilon_0} - \frac{g}{\varepsilon_0} \mathbf{B}\cdot \nabla\phi, \\
    \nabla \cdot \mathbf{B} &= 0, \\
    \nabla \times \mathbf{E} +\frac{\partial \mathbf{B}}{\partial t}&= 0, \\
    \nabla \times \mathbf{B}- \mu_0 \varepsilon_0 \frac{\partial \mathbf{E}}{\partial t} &= \mu_0 \mathbf{J} + g\mu_0\left( \mathbf{B}\cdot \frac{\partial\phi}{\partial t}+\mathbf{E}\times \nabla\phi\right),
\end{align}

The above modification of Maxwell's equations is known as `axion electrodynamics'. The axion field $\phi$ appears on the  right hand side of the sourced Maxwell equations, like a new type of `dark' charge or current density. The axion field only appears with derivatives, i.e. $\partial\phi/\partial t$ or $\nabla\phi$. The axion field always appears multiplied by $\mathbf{E}$ or $\mathbf{B}$. These properties bear out our intuition from above: gradients of the axion field can act as sources for electric and magnetic fields. If you inspect the vector properties of the above equations, you see that the axion always sources an $\mathbf{E}$ (anti)parallel to an applied $\mathbf{B}$ and vice versa. This last fact is in distinction to an electromagnetic wave, in which electric and magnetic fields are perpendicular to one another. Lastly, the magnitude of the axion source term depends on the unknown coupling constant $g$.

Axion electrodynamics has a number of other unusual properties, and is a fascinating subject in and of itself~\cite{Wilczek:1987mv} (for some solution strategies, see Refs.~\cite{Millar:2016cjp,Jeong:2023bqb}). Recently there has been growing interest in something called `axion quasiparticles'. An axion quasiparticle is a system in materials science that has the properties of axion electrodynamics, caused by something called the magnetoelectric effect, but where the field $\phi$ is not a fundamental particle, but instead some collective excitation of the material, such as magnetic order~\cite{Li_2010,Sekine:2020ixs}. In such materials, one could study the physics of axion electrodynamics, and some properties may be useful in microelectronics. So far, materials scientists have fabricated materials with \emph{static} $\phi$, but have yet to realise dynamics. My personal interest in these materials is that, in a wonderful coincidence of physics, the axion quasiparticles might prove useful for detecting axion dark matter~\cite{Schutte-Engel:2021bqm}.

%%%%%%%%%%%%%%%%%%%%%%%%%%%%%%%%%
\section{The search for axions}

\subsection{Astrophysical searches}

Axions can affect numerous astrophysical processes, from the lifetimes of stars, to the large scale structure of the Universe. Reviews can be found in Refs.~\cite{Raffelt:2006cw,Marsh:2015xka,Marsh:2021jmi}. Here we touch on just two possibilities. The two we discuss are particularly important because they do not rely on the assumption that axions are dark matter. The first constraint provides an upper bound on the coupling, $g$, and the second excludes some range of values for the axion mass, $m_a$.

\subsubsection{Axions and the lifetimes of stars}

As was mentioned above, the interaction Hamiltonian in Eq.~\eqref{eqn:em_interaction}, as well as allowing axions to source EM waves in Maxwell's equations, also allows EM waves to source axions. The process is called the Primakoff process, and works like this. Inside a star, atoms are heated up so that they are ionized, which means there are free positively charged nuclei and negatively charged electrons around. These free charges source electric fields by Coulomb's law. There are also a lot of EM waves around: it's hot inside a star. When an EM wave moves into the electric field of a charged particle, then we have the $\mathbf{E}$ in the Hamiltonian from the Coulomb field, and the $\mathbf{B}$ part is provided by the oscillating part of the EM wave. This means that, to minimise the energy, the axion field constantly moves up and down from positive to negative values, as the $\mathbf{B}$ field in the EM wave does. The result is that axion waves are produced. 

To fix up energy conservation, the EM wave has to lose energy. We know from the photoelectric effect that EM waves are actually quantised as photons, with energy $E=h \nu$, and so in the quantum picture the EM wave loses energy in these chunks, creating axions of the same energy. This extra way in which the EM field can interact with nucleons speeds up the evolution of the star. 

This phenomenon, Primakoff production of axions inside stars, has two consequences. The first is that if axions exist, then stars should age a little bit faster than they would otherwise. The rate of stellar ageing depends on the size of the coupling constant, $g$, with larger $g$ leading to more axions produced, and faster ageing. Theories of stellar structure and lifetimes in modern astrophysics are quite advanced. In particular they can be used to accurately predict the famous `Hertzsprung-Russell diagram', that plots stars by their temperature and brightness. In the HR diagram there is a particular region called the horizontal branch, where stars have an almost constant luminosity for a wide variety of temperatures. As stars grow older, they move from the main sequence and onto the horizontal branch as they decrease in temperature approaching the `asymptotic giant branch' which leads eventually to their death. If there is faster ageing of a star caused by axion emission, then the time a star spends on the horizontal branch is decreased. The fact that horizontal branch stars are known to be old allows astronomers to place limits on the amount of extra ageing caused by axions, and thus place an upper limit on the coupling constant $g$~\cite{Raffelt:2006cw}.~\footnote{I thank Maurizio Giannotti for correcting my description of this process.}

The second consequence of axion production in stars is that we can look for the axions produced by the sun. The axions produced have a typical energy determined by the internal temperature of the star, $E=k_{\rm B} T\approx 1\text{ keV}$. These axions travel to Earth. Using the inverse Primkoff process, i.e. converting an axion wave into an oscillating electric field inside a large magnet, turns these axions back into visible X-rays, even if the magnet pointed at the Sun is inside a dark room. This idea is known as an axion `helioscope' and was proposed by Sikivie in 1983~\cite{Sikivie:1983ip}. A number of helioscopes have been built, the biggest being the Cern Axion Solar Telescope~\cite{CAST:2007jps}. They haven't seen anything, and this allows an upper limit on $g$ to be set, which happens to be around the same value you get from lifetimes of horizontal branch stars, so we wouldn't expect to really see anything or stellar evolution would have already shown us there was extra cooling due to axions.

The next step in this game is to try and build a bigger helioscope to test smaller values of $g$. The project is known as the International Axion Observatory~\cite{Armengaud:2014gea,IAXO:2019mpb}, and has the opportunity to test models where $g$ is smaller by around a factor of ten, and could discover the QCD axion if it happens to have a mass between 3 to 100 meV, in certain models~\cite{Dafni:2018tvj,Hoof:2021mld}.

\subsubsection{Axions and black hole spins}
%%%%%%%%%%%%%%%%%%%%%%%
\begin{figure}
    \centering
    \includegraphics[width=0.6\textwidth]{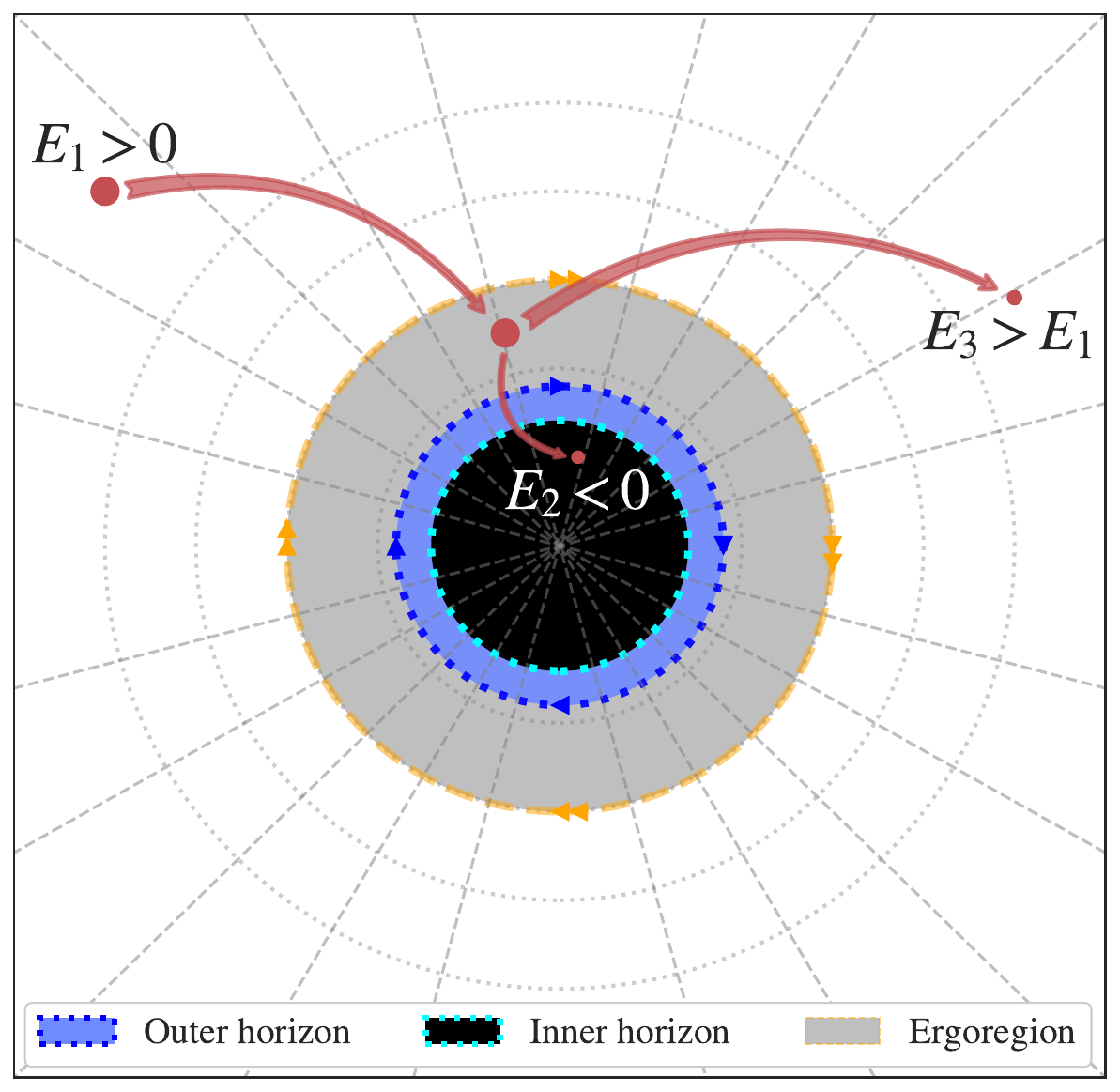}
    \caption{Cartoon of the `Penrose process'. A particle falling into the ergoregion of a rotating black hole can split in two, and leave the ergoregion with more energy than it went in with. This process extracts angular momentum from the black hole. If the process can repeat, the rate grows exponentially, becoming `superradiant'. (Reproduced from Ref.~\cite{Mehta:2021pwf})}
    \label{fig:penrose}
\end{figure}
%%%%%%%%%%%%%%%%%%%%%%%%%

Things famously fall into black holes and don't escape. When things fall in, the black hole mass will increase. If the things falling in also have angular momentum, i.e. falling in on a curved orbit, then the black hole also gains that angular momentum. The only thing that can decrease a black hole mass is Hawking radiation, a very weak quantum process that would take many times the age of the Universe to significantly reduce the mass of an astrophysical black hole. Axions, or in fact any light boson, can change this story dramatically.

Spinning black holes have a region known as the ergoregion around them, where spacetime itself co-rotates, and all objects must follow the spin of the black hole. In 1971, Penrose and Floyd showed~\cite{Penrose:1971uk} that if a particle falls into this region, then splits in two, it must gain energy as the new lighter particle rotates faster. One of the pair falls into the black hole, and the other gains energy is ejected from the ergoregion. Energy conservation is balanced by the black hole losing angular momentum, and spinning a little bit slower. This process is illustrated in Fig.~\ref{fig:penrose}. If this process happens over and over by trapping the outgoing particle, then the rate grows exponentially becoming `superradiant'.

Axion waves around a black hole provide for this scenario naturally if the quantum wavelength of the axion wave, the Compton wavelength, $\lambda_{\rm C}=h/m_a$ is the same size as the ergoregion itself, $r_{\rm Ergo}\approx G_{\rm N} M_{\rm BH}$. In this case, the orbitals of the axion around the black hole, like the orbitals of an electron in hydrogen, provide the mechnism for the process to repeat, by trapping the axion, while the orbitals themselves overlap the horizon of the black hole, allowing for the slow trickle of axions into the black hole necessary for the Penrose process. The mathematics of this process are reviewed in Ref.~\cite{Brito:2015oca}.

What this process means is that if an axion exists with $h/m_a=G_{\rm N} M_{\rm BH}$ then we should expect black holes of mass $M_{\rm BH}$ to only be spinning slowly. On the other hand, the observation of highly spinning black holes of mass $M_{\rm BH}$ excludes the existence of axions of mass $m_a=G_{\rm N} M_{\rm BH}/h$~\cite{Arvanitaki:2010sy}. There are many observations of spinning black holes, and these in fact do exclude the axion from having certain values of the mass, in particular the excluded ranges are (e.g. Ref.~\cite{Mehta:2021pwf}):
\begin{align}
10^{-13}\text{ eV}\lesssim &m_a \lesssim 10^{-11}\text{ eV} \quad \text{(stellar mass black holes)}\, , \\
10^{-19}\text{ eV}\lesssim &m_a \lesssim 10^{-17}\text{ eV} \quad \text{(supermassive mass black holes)}\, .
\end{align}
The exclusions ignore axion self-interactions, which if large quench the process. Thanks to black hole superradiance, we know where not to look for axion DM. In future, large statistical surveys of black hole spin distributions could also reveal the existence of axions~\cite{Arvanitaki:2016qwi}.

\subsection{Microwave cavity `haloscope': the archetype axion DM search}

%%%%%%%%%%%%%%%%%%%%%%%
\begin{figure}
    \centering
    \includegraphics[width=0.6\textwidth]{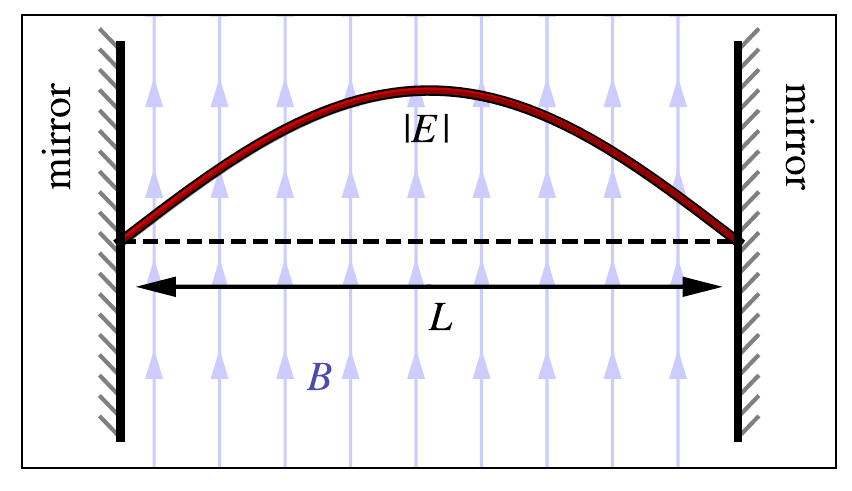}
    \caption{Sinusoidal shape of the fundamental mode of the electric field, $E$, when confined between two conducting mirrors. In an axion haloscope, a $B$ field is applied parallel to the surface of the conductor throughout the region of vacuum between the plates. (Reproduced from Ref.~\cite{DM_book}, published by Princeton University Press, forthcoming)}
    \label{fig:haloscope}
\end{figure}
%%%%%%%%%%%%%%%%%%%%%%%%%

How can we detect axion DM waves on Earth? As we have seen already, the solution to Eq.~\eqref{eqn:KG_homo} is an oscillating classical wave, with frequency $\nu = m_a c^2/h$. From the modified Maxwell equations of axion electrodynamics, we can see that in the presence of an external magnetic field, this oscillating axion wave will source an oscillating electric field, which in principle we could detect. The magnitude the induced electric electric field is extremely tiny, even for very large magnetic fields in the 10's of T.

However, the induced electric field can be enhanced by exploiting \emph{resonance}, in an idea also put forward by Sikivie in 1983~\cite{Sikivie:1983ip}. In a conductor, the electric field must be zero, and acts like a mirror for electric fields. If we place two conducting surfaces parallel to each other, with a vacuum in between, the electric field must be zero at both conductors, and between them can take any configuration with these boundary conditions. If the distance between the plates is $L$, and the first plate is at $x=0$, then such an electric field can be written as a sum of sine waves, known as modes:
\be
E = \sum_{n=1}^{\infty} E_n \sin \left( \frac{n\pi x}{L}\right)\, . \label{eqn:e-field-expansion}
\ee
The first, or fundamental mode, in such a set up is shown in Fig.~\ref{fig:haloscope}. This is the principle of a microwave cavity.

If axion DM waves are present, and the microwave cavity is exposed to a magnetic field parallel to the walls, then the field induced by the axion DM must obey the cavity boundary conditions, and can be written as in Eq.~\eqref{eqn:e-field-expansion}. Now, because Maxwell's equations obey relativity, then the electric field time oscillations must be equivalent to its spatial oscillations, up to a factor of $c$, i.e. for the fundamental mode we must have:
\be
E = E_1 \sin \left( \frac{\pi x}{L}\right)\sin \left( \frac{\pi ct}{L}\right)\, ,
\ee
so that the mode has a `natural frequency' $\nu = \pi c/L$. If the natural frequency of the mode matches the axion DM frequency, i.e if $\pi c/L = m_a c^2/h$, then the amplitude $E_1$ will be resonantly enhanced. We do not know the axion mass, and thus do not know its frequency, but in this toy model we could change the distance $L$ between the plates to change the frequency of the fundamental mode and `tune' the cavity until we reach the right value for resonance. If the resonantly enhanced electric field, $E_1$, is large enough, then it could be detected by a microwave antenna placed between the plates, and we would see a signal appear as if out of nowhere when the frequency is tuned just right, thus detecting axion DM, and simultaneously measuring its mass from the resonant frequency.~\footnote{For a Contemporary Physics article on related and parallel topic, the use of atom interferometers to search for ultralight scalar dark matter, see Ref.~\cite{Buchmueller:2023nll}.} 

To make such an idea realistic requires a few complications. Fortunately, microwave cavities have many applications throughout physics and engineering, and their design and optimisation is well understood. Firstly, no conductor is perfect, and electric fields are not exactly zero on the boundaries. This leads to losses of energy that damp the resonance, and mean that power amplification by such means can be no larger than the `quality factor' of the cavity, $Q$, which for copper can be as large as $10^5$ or so, or for a superconductor can be as large as $Q=10^{10}$. Next, we should worry about how large the applied magnetic field can be, and how large the cavity can be. Commercial magnets can easily get in the 10's of Tesla with volumes in the hundred litre range. 

\begin{figure}
\centering
\subfloat[3D representation of axion photon conversion in a cavity.]{%
\resizebox*{7cm}{!}{\includegraphics{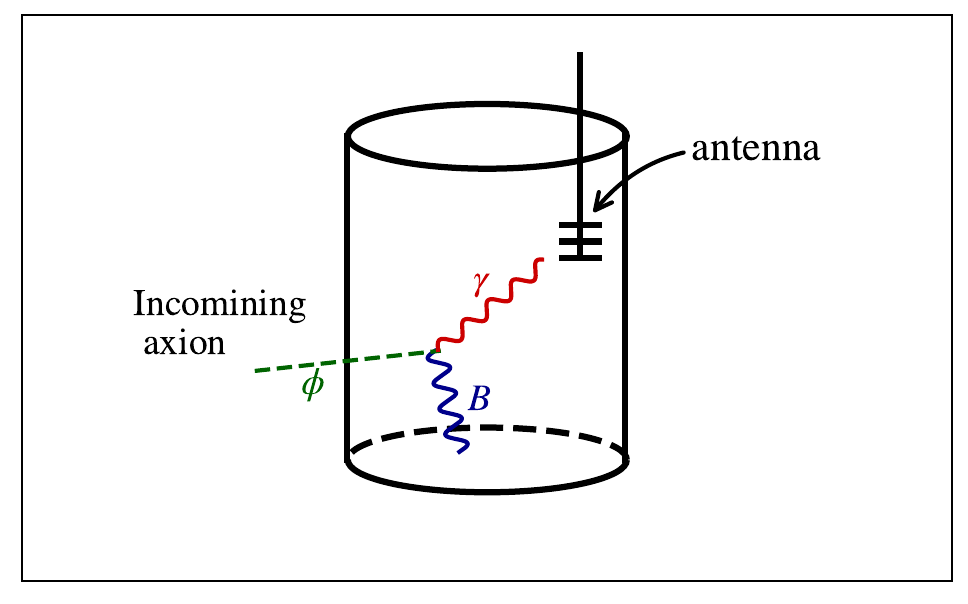}}}\hspace{5pt}
\subfloat[Top down view of a cavity showing tuning rods.]{%
\resizebox*{7cm}{!}{\includegraphics{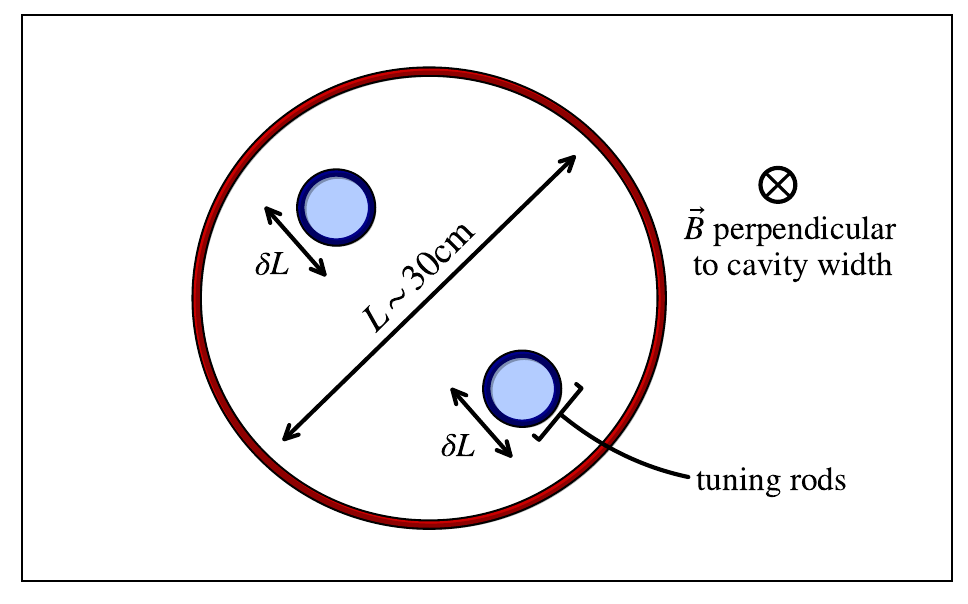}}}
\caption{Sketch of an axion haloscope. (Reproduced from Ref.~\cite{DM_book}, published by Princeton University Press, forthcoming)} \label{fig:3D_2D_cavity}
\end{figure}

There should be no signal in the absence of axion DM, which means isolating the experiment at very low temperature. A commercial dilution refrigerator can cool such an apparatus to temperatures below 1 Kelvin, -272 Celsius. A real microwave cavity is not a simple pair of parallel plates, but likely a cylinder, and its modes are slightly more complex in shape, and can be tuned for example by placing a rod in the cavity. Lastly, even with all of these advances, the signal is still tiny, with axion DM producing a power in the cavity in the range of $10^{-22}\text{ W}$, which is challenging to detect even with the most sensitive antennas and amplifiers. A slightly more realistic sketch of an axion haloscope is shown in Fig.~\ref{fig:3D_2D_cavity}.
%%%%%%%%%%%%%%%%%%%%%%%%%%%
%%%%%%%%%%%%%%%%%%%%%%%%%%%%%%%

The first experiment using this design was carried out in 1987~\cite{DePanfilis}.~\footnote{In a remarkable coincidence, some of the first searches for another dark matter candidate, the supersymmetric weakly interacting massive particle (WIMP), were also carried out in this year~\cite{Ahlen:1987mn}. The theory of WIMP DM production, like that of axions, was also developed in 1983~\cite{Weinberg:1982tp,Ellis:1983wd}, after major theoretical breakthroughs in 1981~\cite{Dimopoulos:1981zb}. The futures of these two models, WIMPs and axions, were very different though, with WIMPs very much in the ascendancy throughout the 1990's and early 2000's. This was due, in part, to technology: WIMP DM direct searches developed sensitivity rapidly, and indirect searches piggybacked off the Higgs search at Cern. Axion searches were much more limited by technology, and ideas, until the 2010's and later. Now, the fortunes of axions and WIMPs have largely reversed.} The modern leaders in this field are the Axion Dark Matter eXperiment (ADMX), based in Seattle, Washington, who have been able to push the sensitivity of the technique far enough to test the QCD axion DM model across a range of particle mass~\cite{ADMX:2021nhd}:
\be
2.7\, \mu\text{eV}\lesssim m_a\lesssim 4.2 \,\mu\text{eV} \quad \text{(ADMX exclusion of QCD axion)} \, .\label{eqn:ADMX_DFSZ_range}
\ee
ADMX has found nothing, and has excluded the benchmark historical axion models from composing all of the local DM density.~\footnote{There are some subtleties if the axion clumps into `miniclusters', but this is not expected to be relevant at the mass scale probed by ADMX, see Refs.~\cite{Hogan:1988mp,Ellis:2020gtq,Kavanagh:2020gcy,Hoof:2021jft,Eggemeier:2022hqa}} This is a technological triumph and the leading light of axion searches at present. 

\subsection{The global axion experimental explosion}
%%%%%%%%%%%%%%%%%%%%%%%
\begin{figure}
    \centering
    \includegraphics[width=0.6\textwidth]{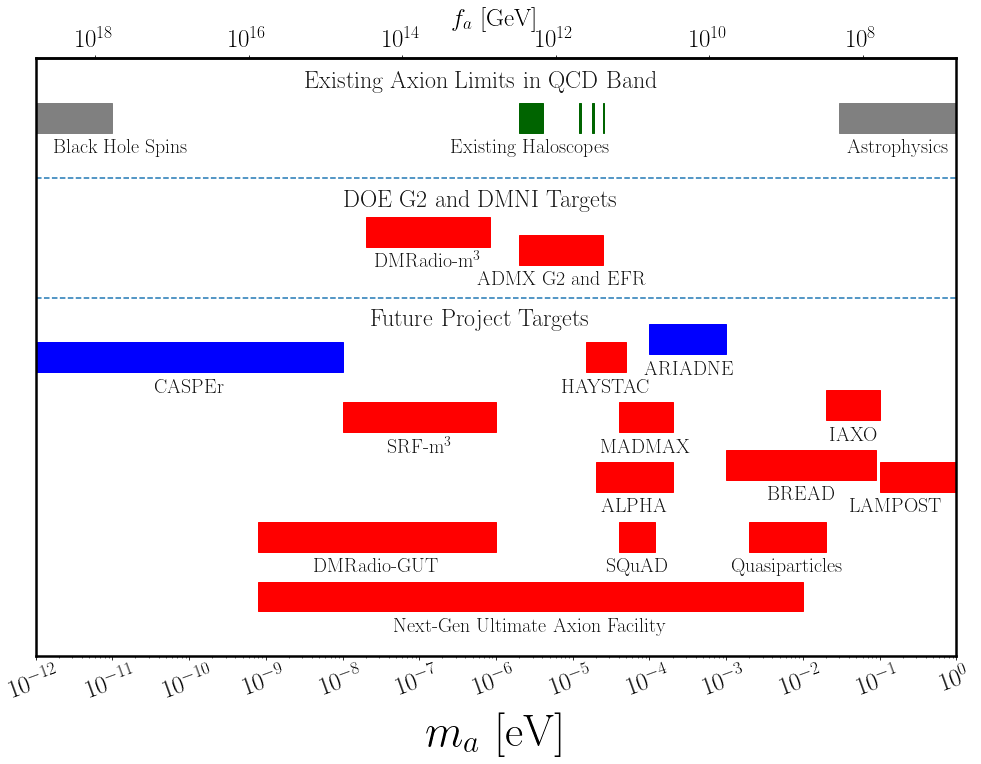}
    \caption{Top row: currently excluded range of masses for the QCD axion. Grey shows upper and lower limits from astrophysics. Black hole spins have been discussed here, while the other astrophysical upper limit comes from supernova cooling. The green region is that excluded by ADMX (Eq.~\ref{eqn:ADMX_DFSZ_range}) and some other currently operating haloscopes. Red and blue regions represent target ranges of new technology planned for the coming decades, which may allow to search the whole allowed range for the QCD axion. (Reproduced from Ref.~\cite{Adams:2022pbo})}
    \label{fig:snowmass}
\end{figure}
%%%%%%%%%%%%%%%%%%%%%%%%%

The mass range probed by ADMX, Eq.~\eqref{eqn:ADMX_DFSZ_range}, is narrow and does not cover anywhere near the entire allowed range for the QCD axion mass:
\be
10^{-11}\text{ eV}\lesssim m_a \lesssim 10^{-2}\text{ eV} \quad \text{(allowed range of QCD axion)}\, .
\ee
The narrowness of the ADMX search range is due to the inherent resonant nature of the haloscope method, and limitations in physical size (both large and small) of microwave cavities with large volume fundamental mode, cooled and subjected to large magnetic fields. 

\begin{figure}
    \centering
    \includegraphics[width=\textwidth]{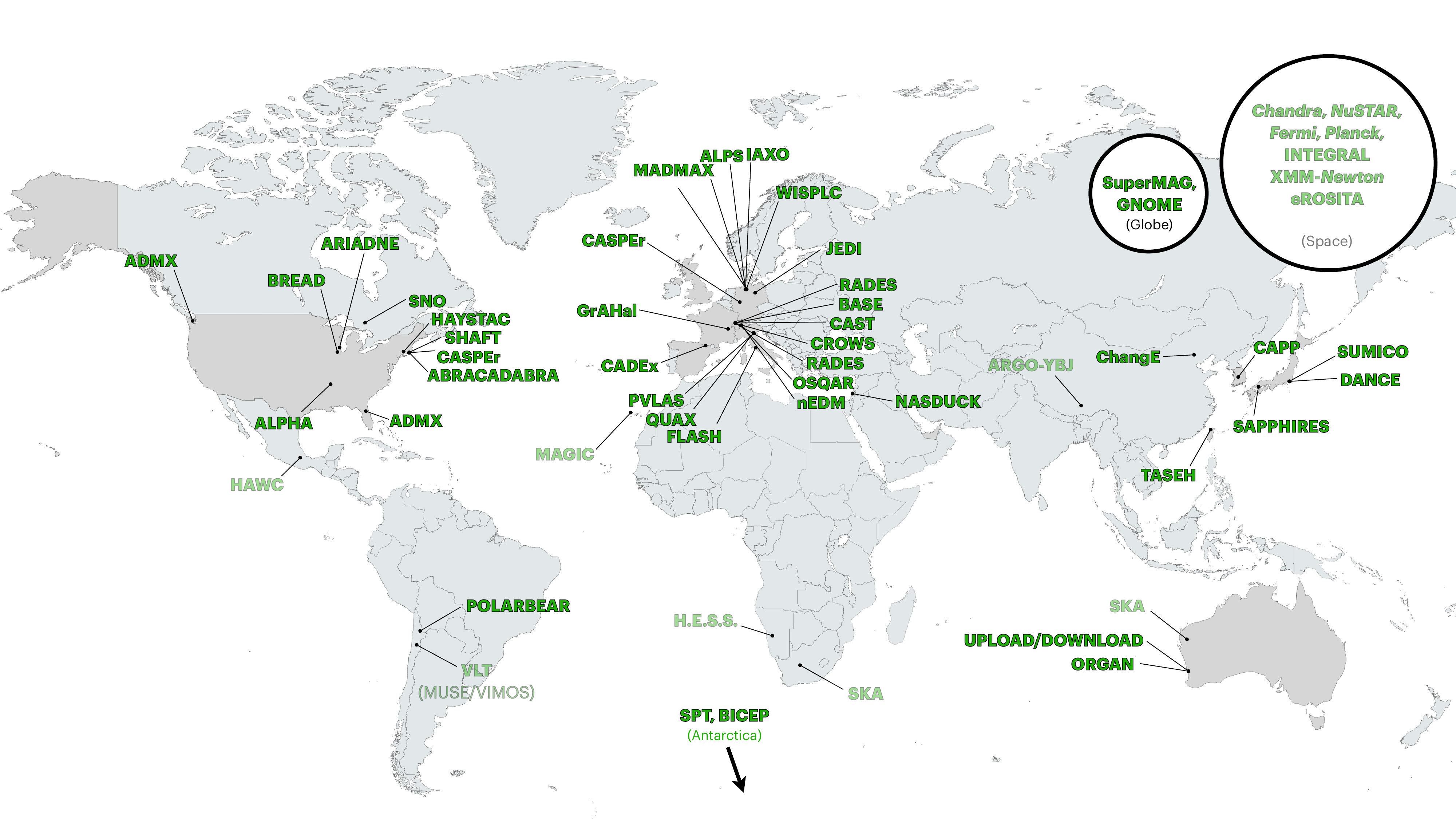}
    \caption{Just a small selection of existing and planned axion experiments around the world (modified from https://cajohare.github.io/AxionLimits/).}
    
    %(map created with Google Maps).
    \label{fig:map}
\end{figure}

The future of axion experiments is to widen the search range using new technology. The current status of the field is reviewed in the white paper Ref.~\cite{Adams:2022pbo}, and summarised in Fig.~\ref{fig:snowmass} in terms of the axion mass search range (with a focus on US developments). The global reach of the modern program is illustrated in Fig.~\ref{fig:map}. We end this article by briefly discussing the exciting global future of the hunt for axions.

A microwave cavity haloscope like ADMX based on Sikivie's original design searches for the axion like tuning a radio. We move to one frequency, listen for the station until we are satisfied even a weak signal is not there, then move on to the next frequency. Steps have to be narrow, because the axion DM station occupies a narrow region in frequency space with bandwidth $\Delta \nu/\nu\approx 10^{-6}$. Furthermore, the frequency is intrinsically limited to searches around 1 GHz by the microwave nature of the technology. Nonetheless, there is still a large frequency range open to this technology, and other haloscopes are being constructed in countries around the world, each working at a different frequency range, and hoping that they will be the ones lucky enough to hit the jackpot of discovering the axion. The nature of the searches makes the experiments not in direct competition: if one experiment hits the jackpot, it was nature that chose the frequency. This makes for a very open and collegial field. Cavity technology has developed rapidly in recent years, lead by innovations from around the world, for example the group surrounding the ORGAN experiment in Australia~\cite{McAllister:2017lkb}, and the Centre for Axion Precision Physics (CAPP) in South Korea~\cite{Semertzidis:2019gkj}.

Searching for axions at higher and lower frequencies than microwave haloscopes can reach, however, has required the emergence of new ideas. At high and low frequency, the trick has been to break the relatioship between the resonant frequency enhancement and the size of the cavity, which we saw above requires $\pi c/L = m_a c^2/h$.
%%%%%%%%%%%%%%%%%%%%%%
%%%%%%%%%%%%%%%%%%%%%%

At high frequency, there are two leading ideas under development, which build on the experience of ADMX. The first takes the cavity idea and instead constructs a kind of open resonator, with the field mode shapes modified by dielectric materials~\cite{Caldwell:2016dcw,Millar:2016cjp}. Another similar idea uses conducting wires to create a tuneable `metamaterial'~\cite{Lawson:2019brd}. Prototypes of both of these ideas have been created, and plans are in place to scale the prototypes to full axion DM searches. For the dielectric case, the there is `MADMAX', which is planned to operate in Europe at DESY and Cern, and also `DALI' under development in Spain~\cite{DeMiguel:2020rpn}. In the metamaterial case, the leading proposal is `ALPHA', which is being developed largely in the USA. The quasiparticles idea mentioned above~\cite{Schutte-Engel:2021bqm} also falls in this broad category at high frequency. An alternative approach at high frequency is to give up on resonance and instead use a large reflector to collect axions from a wide magnetized source region, with the advantage that such a search is broadband~\cite{Horns:2012jf}. Planned and operating experiments using this technology include `BRASS' in Germany~\cite{Bajjali:2023uis}, and `BREAD' in the USA~\cite{BREAD:2021tpx}. Together, these technologies may reach up to the highest axion masses corresponding to THz frequencies.

At low frequency, a slightly different approach is taken, with the most promising set of experiments being based on the ideas developed in Refs.~\cite{Chaudhuri:2014dla,Kahn:2016aff}. In this proposed `DM-Radio'~\cite{DMRadio:2022jfv,DMRadio:2022pkf}, and related designs, the axion interaction with a magnetic field produces a current in an LC circuit resonator, which is then detected with a pickup loop. The frequency sensitivity of this technique extends down as low as kHz.

Axions can also interact with ordinary matter via different means than axion electrodynamics, for example coupling to electrons or nuclei directly. Searching for axions in this way involves resonances, either nuclear magnetic resonance (NMR), or electron spin resonance, which are present in certain types of material. The NMR searches use precisely the technology developed already in that advanced field, and apply it to search for axions, where they appear as a tiny anomalous magnetic field. This scheme of experiments is called CASPEr, and operates in Germany and the USA~\cite{Budker:2013hfa,JacksonKimball:2017elr}. On the side of the electron interaction, the experiment QUAX was developed in Italy~\cite{Barbieri:2016vwg,QUAX:2020adt} (early ideas in this direction were developed in Refs.~\cite{Barbieri:1985cp,Kakhidze:1990in}). These searches for axions using material resonances and different couplings between the axion and ordinary particles will be of key importance in a `post-discovery' world, with in particular CASPEr being able to distinguish the QCD axion from a more general ALP, and QUAX being able to disentangle between the two key types of QCD axion model.

\section{Closing remarks}

In this brief review we have been on a journey from the smallest scales, with axions in extra dimensions, to the largest scales, with axions dictating the evolution of the Universe. We have met physics from General Relativity, to the design of microwave cavities. This is one reason to be excited by axions: they give the physicist studying them ample opportunity to learn across a wide range of specialisms. But the much more pressing reason to be excited by axions is the massive growth in experiments searching for axions. Many of the experiments mentioned above are only in the prototype stage, and will reach design sensitivity in the next 10-20 years. This means that the coming decades could well see the discovery of the QCD axion, if it exists. If axions are discovered, this could shed light on the mystery of dark matter, the strong-$CP$ problem, and even the topology of extra dimensions. There could not be a more exciting time to come.

\section*{Funding}

I am supported by an Ernest Rutherford Fellowship from the Science and Technologies Facilities Council (ST/T004037/1), and by a Leverhulme Research Project (RPG-2022-145).

\section*{Nomenclature/Notation}

The gradient operator in three dimensions is $\nabla$, and in this context $\times$ is the vector cross product. The speed of light is $c$, Planck's constant is $h$. Particle masses are quoted in units of electronvolts, eV, where $1\text{ eV}= 1.78\times 10^{-36}\text{ kg}$, and an atom of hydrogen is approximately $10^9\text{ eV}$. Particle physicists often used units where $\hbar=c=1$, and while I have tried my best to restore these factors, as well as those of $\epsilon_0$ and $\mu_0$, I cannot guarantee I caught every one.

\bibliographystyle{tfnlm}
\bibliography{Axion}

\appendix

\end{document}